\documentclass{apex}

\title{Generation of Nano-Catalyst Particles by Spinodal Nano-Decomposition 
in Perovskite}

\author{Hidetoshi Kizaki\thanks{E-mail address: hkizaki@aquarius.mp.es.osaka-u.ac.jp}, Koichi Kusakabe, Soichiro Nogami, and Hiroshi Katayama-Yoshida}
\inst{Graduate School of Engineering Science, Osaka University, Toyonaka, Osaka 567-8531, Japan}
\abst{
A new mechanism of nano-catalyst generation 
based on the spinodal nano-decomposition 
in self-regenerating perovskite catalysts 
for automotive-emissions control is proposed. 
To demonstrate existence of the spinodal nano-decomposition 
in real perovskite catalysts, we performed 
first-principles calculations to evaluate 
the free energy of La(Fe$_{1-x}$Pd$_x$)O$_3$ and 
La(Fe$_{1-x}$Rh$_x$)O$_3$. 
The result indicates appearance of 
a spinodal region in the phase diagram of each material. 
Formation of nano-catalyst particles in the perovskite host matrix 
is crucial for the self-regeneration of perovskite catalyst. 
Based on the spinodal nano-decomposition model, 
possible materials are designed for new three-way catalyst 
with no contents of precious metal. 
}

\begin{document}

\maketitle


Self-regenerating Pd-, Rh-, and Pt-doped perovskite catalysts 
for automotive-emissions control 
are attracting much interest due to their unique 
functionalities\cite{Tanaka-2001,Nishihata-2002,Tanaka-2004,Nishihata-2005,Uenishi-2005b,Tanaka-2006b,Tanaka-2006d,Tanaka-2006c,Tanaka-2006a,Tan-2006,Taniguchi-2007}. 
In a conventional catalyst made of fine precious-metal particles supported on 
a solid like an almina, agglomeration of the metal particles is inevitable 
because of high temperature conditions in the redox environment. 
This is the very reason for deterioration of automotive three-way catalysts. 
In the self-regenerating perovskite catalysts, interestingly, the deterioration 
is strongly suppressed. Therefore, consumption of the precious metal 
is greatly reduced, providing a highly efficient solution for supply problems. 
The reason for non-deterioration is thought to be reformation of 
a precious-metal doped perovskite lattice in the NO$_x$-reduction 
environment from segregated nano-particles of precious-metal 
created in the CO- and C$_x$H-oxidation environment. 
In this regeneration model, precious-metal atoms are assumed to 
move into and out of the perovskite host matrixes. 
Growth of precious-metal grains is suppressed due to this repeated 
motion of precious metal atoms between a solid solution 
and metallic nano-particles during three-way catalytic reactions. 

In some doped perovskite structures, it is known that diffusion of 
oxygen happens rather frequently. At the same time, we may expect 
motion of metal atoms via a process exchanging metallic atoms. 
The process would be enhanced, if oxygen vacancies are created in 
the perovskite host crystal. This scenario might support 
the above model of self-regenerating perovskite catalysts. 
For realization of high efficiency in the three-way catalytic 
functions with no deterioration, 
however, we should consider another model of the 
self-regenerating catalyst. 
The new model should explain some mysteries known experimentally 
in the perovskite catalyst. 
The self-regeneration can happen, only when the host perovskite 
lattice structure maintains its essential structure. 
We need to know why the redox reaction keeps its cycle 
in a redox environment changing in a frequency of about 
a few hertz ($1\sim 4$ Hz). Thus we need to understand motion of 
precious metal atoms in a rather hard oxide. 
Once melting happens in some process of catalytic reaction, 
we again cannot escape from agglomeration of metallic atoms 
leading to deterioration. 
If these problems are solved, we can hope to design 
much efficient catalysts from the knowledge of nano-scale 
structure of solid catalysts. 

To answer these questions, we propose a new model of 
the self-regenerating perovskite catalyst based on the 
spinodal nano-decomposition 
(SND).\cite{Sato2005,Fukushima2006,Sato2007a,Sato2007b,Sato2007c} 
In several covalent and ionic materials, 
it is known that growth of nano-scale structures 
happens in a thermal non-equilibrium synthesis process of the materials 
due to SND. 
One of the most remarkable example is SND 
appearing in diluted magnetic semiconductors. 
The mechanism of SND 
was first clarified theoretically.\cite{Sato2005} 
Later, many experimental evidences supporting 
appearance of SND were found.\cite{Gu2005,Jamet2006} 
Actually, enhancement of the blocking temperature 
appears in the super-paramagnetic substance, 
in which ferromagnetic hysteretic behavior in magnetization 
process happens. Novel functional device structures 
and new functional materials are proposed based on 
SND.\cite{Katayama-Yoshida2007a,Katayama-Yoshida2007b} 
In the perovskite catalysts, 
due to difference in the solubility of metallic atoms in the host matrix, 
there can happen SND. 
In LaFeO$_3$, for example, solubility of precious metal is low 
and it should be different for Pd, Rh, and Pt. 
Once the phenomenon occurs, 
the above puzzling problems are naturally solved as shown in this paper. 

Our purpose of the present study is thus summarized 
in the following steps. 
At first, 
to support occurrence of SND, 
we will give the first-principles simulation using 
the Korringa-Kohn-Rostoker method 
with the coherent-potential approximation (KKR-CPA).\cite{Akai-1993} 
The numerical data supports SND 
at a reasonable chemical composition and at a plausible temperature. 
We will thus proceed to consider a physical picture 
concluded by SND, 
explaining the SND model in detail. 
Phenomenological solutions for the problems 
in perovskite catalysts are given. 
Functionality of the nano-scale catalyst is a key point, 
which gives a natural explanation of the self-regenerating catalysts. 
Several candidate materials for the new three-way catalysts are proposed 
based on the materials design using SND. 

We consider La(Fe$_{1-x}$Pd$_x$)O$_3$ as 
the typical perovskite catalyst. 
To explore material dependence, 
La(Fe$_{1-x}$Rh$_x$)O$_3$ is also examined. 
In these examples, the host material is LaFeO$_3$. 
The position of doped precious metal is known to be at 
the B site of the perovskite 
structure.\cite{Nishihata-2002,Nishihata-2004,Tanaka-2006d} 

In covalent ionic crystals, 
the spinodal decomposition or the phase 
separation often happens depending on the composition and the temperature. 
For La(Fe$_{1-x}$Pd$_x$)O$_3$, depending on 
synthesis processes, the sample can microscopically 
decompose into LaFeO$_3$ and LaPdO$_3$. 
When SND happens, 
spatial inhomogeneity in distribution of Pd atoms occurs 
but the crystal structure itself is maintained. 
X-ray diffraction will see average crystallinity only, 
if the size of each nano-scale structure is less than 1 nm. 
To characterize the distribution and 
to discriminate SND from the phase separation, 
we are required to utilize electron energy loss spectroscopy (EELS) 
or energy dispersive X-ray spectroscopy (EDXS).\cite{Gu2005,Jamet2006} 
To show a support for this message, 
we have performed estimation of the mixing energy 
of LaFeO$_3$ and LaPdO$_3$, or that of LaPdO$_3$ and LaRhO$_3$. 
The mixing energy $\Delta E$ is given 
for La(Fe$_{1-x}$Pd$_x$)O$_3$ as 
$\Delta E(x) = 
E({\rm La}({\rm Fe}_{1-x}{\rm Pd}_x){\rm O}_3)
-\{(1-x)E({\rm La}{\rm Fe}{\rm O}_3)
+xE({\rm La}{\rm Pd}{\rm O}_3)\}$, where 
$E({\rm M})$ is the total energy of a material M. 
If $\Delta E$ is positive, the system has a tendency 
toward spinodal decomposition (or phase separation), 
while negative $\Delta E$ 
suggests that the system favors a homogeneous mixing. 
The actual simulation was done utilizing 
the KKR-CPA program package \textsc{Machikaneyama}2002 
developed by H. Akai.\cite{Machikaneyama-2002} 

In Fig. 1 (a) and (b), we show the results for 
La(Fe$_{1-x}$Pd$_x$)O$_3$ and 
La(Fe$_{1-x}$Rh$_x$)O$_3$, respectively. 
The obtained density of states for a 
5\% Pd-doped LaFeO$_3$ system is shown in Fig. 1 (c). 
At zero temperature, both of the results indicate 
tendency to the spinodal decomposition. 
As shown in the figures, the mixing energies show 
strong concavity as a function of Pd or Rh impurity concentration. 
To show the temperature dependence, we introduce 
the mixing entropy to have a free energy 
as $F=\Delta E(x)-TS$ with 
$S=-k_B[x \log x + (1-x) \log (1-x)]$. 
In this estimation, some temperature effects 
including the volume expansion and lattice deformation, 
kinetics of atoms leading to atomic diffusion, defect formation, 
and crack creation, are not considered. 
However, the present test should be the first prior test 
to judge whether mixing happens in equilibrium or not. 
The results shown in Fig. 1 (a) and (b) 
indicate clearly the phase separation 
($\partial^2 F/\partial x^2 > 0$) 
or spinodal decomposition 
($\partial^2 F/\partial x^2 < 0$) 
at a finite temperature plausible for the ordinal 
synthesis process. 
The free energy curves suggest that, 
only above a rather high temperature ($T>3000$ K for Pd ($x=10$\%) 
and $T>2000$ K for Rh ($x=10$\%)), mixing may happen. 

If the synthesis is done in non-equilibrium condition, 
Pd and Rh could be dispersed in the host matrix. 
However, our simulation strongly suggests that 
the solubility of Pd and Rh in LaFeO$_3$ is small and 
SND should occur in the concave region of 
$F$ ($\partial^2 F/\partial x^2 < 0$), 
when calcination is performed. 
Depending on the heat condition and the calcination time, 
the size of segregated precious metal oxide grain 
due to SND 
might be controlled. 

Our simulation results show that the tendency towards 
the spinodal decomposition is less for the Rh-compound 
than the Pd-compound. Actually, the value of 
$\Delta E(x)$ is about 15\% smaller for the Rh-compound 
than that of the Pd-compound. 
Larger mixing energy for La(Fe$_{1-x}$Pd$_x$)O$_3$ indicates that spinodal 
decomposition should be more serious for this material 
than for La(Fe$_{1-x}$Rh$_{x}$)O$_3$. 
Indeed, we can clearly see in Fig. 1 (a) and (b) at T=1000K that 
$\partial^2 F/\partial x^2 < 0$ 
for Pd ($x=5$\%) increasing SND, however, 
$\partial^2 F/\partial x^2 > 0$ 
for Rh ($x=5$\%) indicating no spinodal decomposition.
This result is consistent with the experimental observations. 
The system in La(Fe$_{1-x}$Rh$_{x}$)O$_3$ 
favors a homogeneous mixing, in which Rh atoms 
tends to disperse in LaFeO$_3$ matrix and therefore 
less effective for the catalytic reactions than 
La(Fe$_{1-x}$Pd$_{x}$)O$_3$, as known in the experiment.\cite{Tanaka-2006d} 


We now discuss relevance of SND 
for the realization of the regenerating catalytic function, 
which makes us possible to design new three-way catalysts. 
Before explaining the SND model, let us consider what happens 
if the catalysts are made from uniformly dispersed 
precious metals in the synthesis process and if SND does not occur. 
Then diffusion of the precious metal atoms from the bulk to 
the reacting surfaces are required for efficient redox reactions. 
Then, segregation of precious metal atoms is expected. 
Nano-particles found in XAFS experiments have been 
thought to be created by this process in the uniformly 
distributed model. 
Once the formation of precious metal particles at the 
reaction surface occurs, leaving defective B-sites in the host, 
agglomeration of the metal particles 
and destabilization of the host matrix are inevitable, 
which leads to some difficulty in understanding the self-regeneration. 

When SND happens in 
a synthesis process, nano-catalysts are created in 
the host matrix. (Fig. 2) In the redox reaction, only the 
nano-catalyst particles react with oxygen atoms going out and into 
the nano-catalysts, keeping the host matrix stable. 
The reaction is enhanced due to high concentration of 
precious (or relevant non-precious) metal atoms 
at the reacting surfaces. 
In the SND model, 
thus we can naturally explain stability of 
the catalysts found in the experiment.\cite{Tan-2005} 

The SND model has 
a big advantage to understand the fast catalytic 
activity.\cite{Nishihata-2004,Uenishi-2005a,Uenishi-2008} 
Since the reactive nano-catalyst particles are formed at the 
reaction surfaces, we can expect very fast start-up for 
the catalytic activity. 
If the host matrix is enough stable, we can hope 
high endurance against the heat, since 
the reacting nano-catalysts keep their size in nano-scale. 
In addition, the nano-scale catalyst particles have 
quantum effects enhancing the catalytic activity. 
Catalytic reactions would be affected by 
discretization of electronic energy levels. 
In nano-catalysts embedded in a host matrix, 
there would be 
appearance of different valence states in a reasonable 
energy range around 
the Fermi level, which would be controlled by conditions of 
the gas phase. 

Following the SND model, 
we can readily design many functional nano-catalysts utilized 
in a wide area of application for the perovskite 
catalysts.\cite{Tanaka-2003,Lohmann-2005,Andrews-2005}
An important selection rule is given by solubility 
of relevant metal atoms in a host matrix. 
We should select relevant $3d$ transition elements 
for the reduction or the abandonment of the precious metal. 
Important candidates should be Mn, Fe, Co, Ni, and Cu. 
Consideirng the solubility, possible materials are, 
Ca(Ti$_{1-x}$M$_x$)O$_3$, 
Sr(Ti$_{1-x}$M$_x$)O$_3$, 
Ba(Ti$_{1-x}$M$_x$)O$_3$, 
Ca(Zr$_{1-x}$M$_x$)O$_3$, 
Sr(Zr$_{1-x}$M$_x$)O$_3$, 
and Ba(Zr$_{1-x}$M$_x$)O$_3$ with 
M$=$Mn, Fe, Co, Ni, Cu, and Zn. 
If we inversely consider functions of elements, 
the next structure might be another solution: 
La(Fe$_{1-x}$M$_x$)O$_3$ with M$=$Ti and V. 

\acknowledgement
This work was partially supported by 
the Elements Science and Technology Project, 
Grant-in-Aid for Scientific Research 
from the Japan Society for the Promotion of Science 
and the Ministry of Education, Culture, Sports, Science and 
Technology (MEXT), 
Global Center of Excellence Program by MEXT, 
New Energy and Industrial Technology Development Organization Program, 
and Japan Science and Technology Agency Program.

\newpage

\begin{figure}[htbp]
\begin{center}
\includegraphics[width=0.8\hsize]{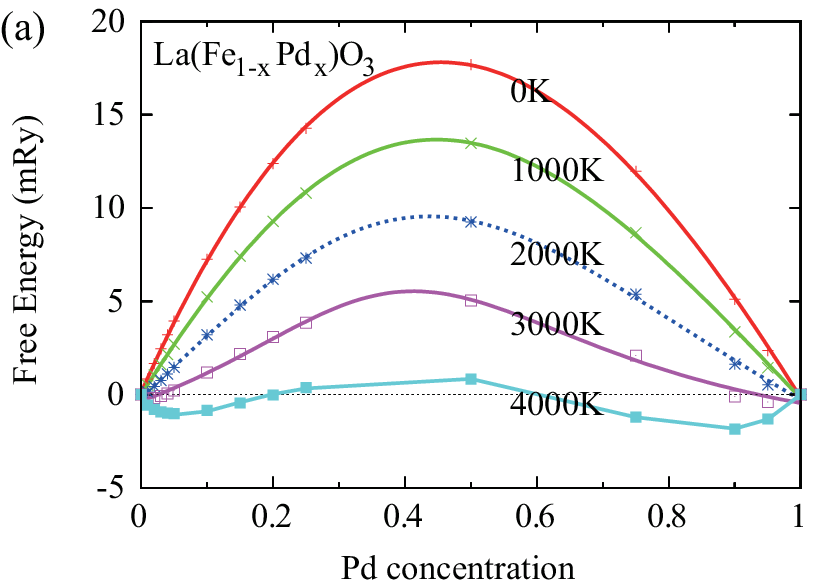}
\includegraphics[width=0.8\hsize]{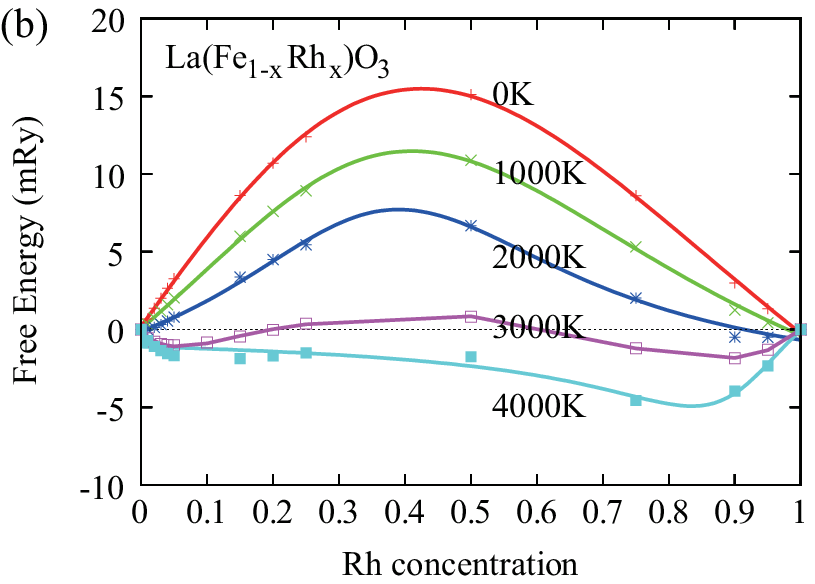}
\end{center}
\end{figure}
\setcounter{figure}{0}
\begin{figure}[htbp]
\begin{center}
\includegraphics[width=0.8\hsize]{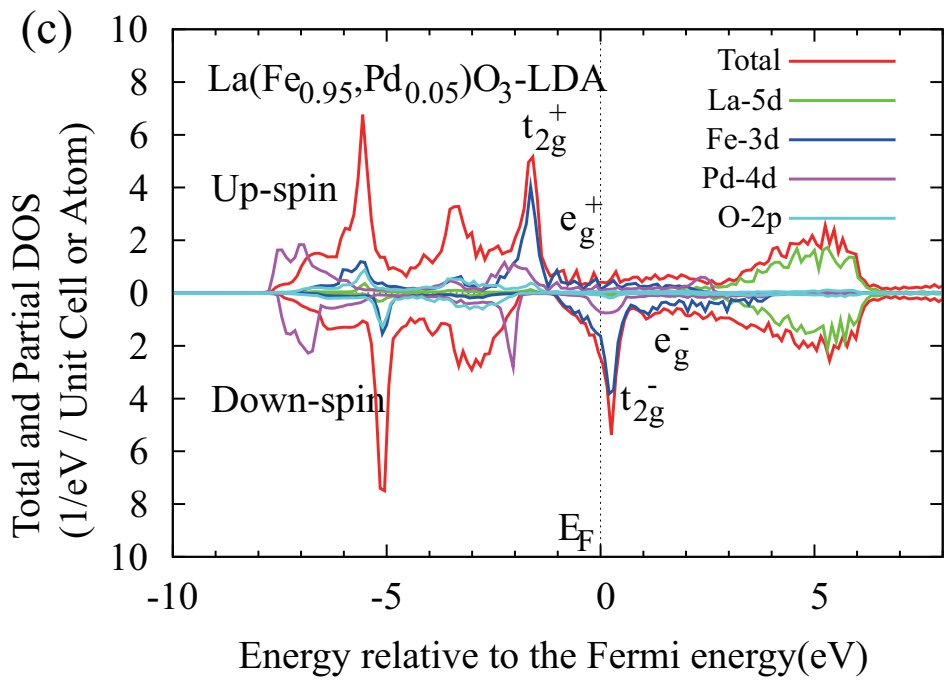}
\caption{
Calculated free energies of 
(a) La(Fe$_{1-x}$Pd$_x$)O$_3$ and (b) La(Fe$_{1-x}$Rh$_x$)O$_3$. 
The free energy is calculated as a function of 
concentration of the precious metal. 
The temperature dependence is obtained by 
introducing the mixing entropy. 
In the panel (c), calculated total 
and partial density of states for La(Fe$_{0.95}$Pd$_{0.05}$)O$_3$ are shown. 
}
\label{fig1}
\end{center}
\end{figure}

\begin{figure}[htbp]
\begin{center}
\includegraphics[width=0.7\hsize]{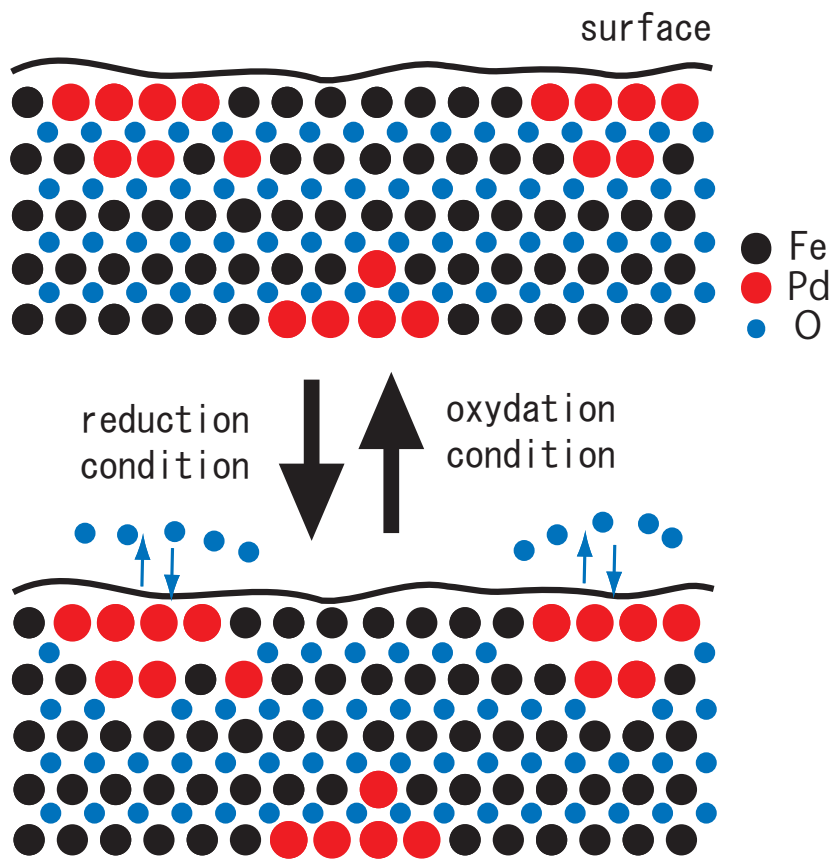}
\caption{
Schematic pictures of the spinodal nano-decomposition model 
of the self-regenerating perovskite catalysts. 
Black, red, and blue atoms are iron, palladium, and oxygen atoms, 
respectively. In the redox reaction, motion of atoms 
shown in each figure should happen. 
La-O bond is strong and stable in the reduction condition. 
Therefore, we do not show La-O in this schematic picture.}
\label{fig1}
\end{center}
\end{figure}

\end{document}